# Neutral naturalness from a holographic $SO(6)/SO(5)$ composite Higgs model

Barry M. Dillon[*]

*Jožef Stefan Institute, Jamova Cesta 39, 1000 Ljubljana, Slovenia*

(Received 2 July 2018; published 6 June 2019)

We study a holographic realization of a composite Higgs model with an $SO(6)/SO(5)$ symmetry breaking coset in which the top sector includes color-neutral twin-partners that reduce the sensitivity of the Higgs mass to the cutoff. Key to this "neutral naturalness" mechanism is a $Z_2$ symmetry that leaves the top Yukawa coupling to the Higgs boson invariant under an exchange of the top quark and twin top quark, but the symmetry structure of the model means that the $Z_2$ symmetry is not present in the gauge boson couplings to the Higgs. Within the calculable framework of holography we construct and study the Higgs potential. We examine the relation between the Higgs mass, top-partner spectra, and the input parameters, finding that the presence of the twin-partners pushes the masses of the lightest colored top-partners up to $\sim 1500$ GeV while the decay constant remains $\lesssim 700$ GeV. Interestingly, no additional $Z_2$ breaking terms are required to reproduce the observed masses of the electroweak gauge bosons, Higgs boson, and top quark.

## I. INTRODUCTION

With the discovery of the Higgs boson at a mass of 125 GeV [1] and the absence of new physics between the electroweak scale and the TeV scale, important questions regarding the naturalness of the Higgs sector must be addressed. Natural scenarios accommodating a light Higgs boson include those with a composite Higgs sector [2]. A central aspect here is that the Higgs degrees of freedom are pseudo-Goldstone bosons of a global symmetry that is spontaneously broken by the condensation of a strongly coupled gauge theory. In recent years these models have received a lot of interest, and many phenomenologically interesting models have been identified [3–6]. The general picture is that a set of fermions with a flavor symmetry $\mathcal{G}$ is coupled to a gauge theory whose coupling grows strong near the TeV scale. The confinement of these fermions into bound states then spontaneously breaks $\mathcal{G}$ to a subgroup $\mathcal{H}$ with the Higgs degrees of freedom being formed from the Goldstone bosons in the $\mathcal{G}/\mathcal{H}$ coset. With $SU(2)_L \times U(1)_Y \subset \mathcal{H}$, coupling the Standard Model (SM) fields to the strong sector explicitly breaks the global symmetries and generates a potential for the Higgs field, which in turn allows for the spontaneous breaking of the electroweak

(EW) gauge symmetry. Assuming that the Yukawa couplings are generated via partial compositeness, the large value for the top Yukawa coupling requires a large mixing between the right-handed (RH) top quark and composite top-partner states. This leads to large contributions to the Higgs potential which can only result in a naturally light Higgs boson when there are light top-partners in the spectrum [7]. Much work on the phenomenology of these light top-partner states has been carried out [8,9]. For a general discussion on the 4D construction of composite Higgs models and little Higgs models see [10–13] and [14], respectively. Recent experimental results put lower bounds on colored top-partner masses in the region $\sim 1$–1.4 TeV [15], thus there is a need for theoretical models which can explain this absence of light top-partner states.

One broad class of models which elegantly evades the collider bounds on top-partners are the "neutral naturalness" models. These scenarios contain additional states that suppress the Higgs potential and allow a light Higgs to exist naturally in the spectrum. However the new states are not charged under the QCD gauge group and thus the bounds from the LHC do not apply in their full generality, allowing for new physics at mass scales closer to the electroweak scale. In the composite Higgs framework there are models of neutral naturalness known as twin Higgs models [16], and in supersymmetry there are models called folded-supersymmetry models [17]. Within the composite twin Higgs [18] landscape recent studies have also highlighted how twin-baryons may be a good dark matter candidate [19]. Other models of neutral naturalness such as the quirky little Higgs [20] or models with completely SM-neutral scalar top-partners [21] have also been studied. Much work on the collider phenomenology [22] of neutral naturalness models has been done in recent years.

[*]barry.dillon@ijs.si

The twin Higgs mechanism fits well within the composite Higgs paradigm as it also posits that the Higgs field is formed of pseudo-Goldstone bosons. Central to this mechanism is a $Z_2$ exchange symmetry between the top quark and a "twin top quark," of which the latter is neutral under SM gauge symmetries. This leads to a $Z_2$ exchange symmetry ($s_h \leftrightarrow c_h$) in the top quark contributions to Higgs potential, resulting in a softening of the potential even in the presence of a large mixing between the top quark and the composite sector. In the minimal models the Higgs and twin Higgs degrees emerge from an $SO(8)/SO(7)$ coset. Both the QCD and electroweak gauge sectors also have a twin counterpart meaning that the $Z_2$ symmetry is exact for the whole SM. A minimal neutral naturalness model based on an $SO(6)/SO(5)$ coset has been proposed [23,24] which contains a similar $Z_2$ symmetry but does not contain a twin electroweak group. The models studied in these papers have slight differences, although the features that allow the neutral naturalness mechanism to work are the same. The authors found that the lightest colored top-partners in these models could easily lay above the bounds set by current LHC analyses, and identified some interesting phenomenological aspects of the models. Many studies on nonminimal composite Higgs models can be found in the literature [25,26], and a lot of work developing UV completions of the composite Higgs scenario has been carried out [27].

In this paper we study a holographic realization of the $SO(6)/SO(5)$ model of neutral naturalness proposed in [23,24], where the models are referred to as the "brother Higgs" or "trigonometric parity" models, respectively. Holography is a well established and indispensable tool used in building calculable effective field theories for strongly coupled gauge theories. These methods have their origins in the AdS/CFT correspondence [28] and became hugely popular in the beyond-the-Standard Model (BSM) model-building community after the introduction of the Randall-Sundrum (RS) models [29,30]. Much work has been done in studying the holographic correspondence between the RS models and strongly interacting field theories [31]. For composite Higgs scenarios there are elegant holographic formulations of partial compositeness [32], through which the SM fermions couple to the strong sector, and of the spontaneous breakdown of the global symmetry, through which the composite Higgs degrees of freedom arise [33]. The one-loop Higgs potential calculated in holography is automatically finite due to 5D locality, negating the need to impose sum rules on the model parameters. Many applications of holography to composite Higgs models have been studied [34], and it has been shown that the 5D volume plays a significant role in determining the relationship between the Higgs mass, the

Higgs vacuum expectation value, and the top mass [35]. A holographic description of the $SO(8)/SO(7)$ twin Higgs model was presented in [36,37], where it was found that realistic electroweak symmetry breaking (EWSB) can take place through the introduction of additional $Z_2$ breaking terms in the Higgs potential. One advantage of the model studied in this paper is that it already includes $Z_2$ breaking terms due to the absence of twin electroweak gauge bosons, and we show that these contributions are enough to ensure realistic EWSB.

Although there is a proposed duality between holographic models and strongly coupled gauge theories, in the holographic $SO(6)/SO(5)$ model the gauge and global symmetries are not directly related to the fermionic content of a strongly coupled gauge theory, as they simply arise through the gauge symmetries in the 5D model. Therefore in our calculations we refer only to these 5D gauge symmetries when discussing the symmetry breaking patterns. For discussion on fermionic UV completions of the $SO(6)/SO(5)$ composite Higgs model presented here we refer the reader to [24].

The paper is outlined as follows. Section II presents the details of the model, beginning with an overview of what we want to achieve, and then outlining the 5D holographic model which does so. In Sec. II A 1 we provide the details on the gauge symmetries in the holographic model. The holographic description of the quark sector is outlined in Sec. II A 2, where the origin of the $Z_2$ symmetry in the Yukawa couplings becomes apparent. In Sec. III we study the Higgs potential of the model, define the form factors as a function of the 5D input parameters, and highlight important features that arise due to the $Z_2$ symmetry. Section III A presents the results of a numerical scan over the parameter space, and discusses the predicted top-partner spectra and the dependence on the parameters of the Higgs potential.

## II. THE MODEL

We start with a brief review of the neutral naturalness mechanism outlined in [23,24], and refer the reader to these papers for a more in depth discussion. We work in the composite Higgs framework with the compositeness mass scale assumed to be above a TeV. The strong sector has a global symmetry $\mathcal{G} = SO(6)$ which is spontaneously broken to $\mathcal{H} = SO(5)$ at the compositeness scale. The Higgs doublet ($H$) and a real singlet ($\eta$) emerge as Goldstone bosons in the $SO(6)/SO(5)$ coset. This is the minimal coset that contains an internal parity leading to the required $Z_2$ exchange symmetry between the top and twin top in the Yukawa sector. The electroweak gauge fields are gauged from the $SU(2)_L$ subgroup of $SO(4)$ and the QCD $SU(3)_c$ gauge group is introduced externally to the global symmetries of the strong sector. In a nonlinear sigma description the Goldstone bosons can be written in an $SO(6)$ vector as

$$\Sigma = \frac{\sin \frac{\Pi}{f_\pi}}{\Pi} \left( \pi_1, \pi_2, \pi_3, \pi_4, \pi_5, \cot \frac{\Pi}{f_\pi} \right) \qquad (2.1)$$

with $\Pi = \sqrt{\pi_a \pi_a}$ and $f_\pi$ being the decay constant of the Higgs field. The Higgs doublet is formed from $\pi_1, ..., \pi_4$ and $\pi_5 = \eta$ is the Goldstone boson of the spontaneously broken $U(1)_\eta$ subgroup of $SO(6)$ (the $SO(6)$ generators are given in Appendix A). In addition to the SM gauge symmetries we also introduce a twin QCD, $SU(3)_{\tilde{c}}$, and gauge the $U(1)_\eta$ subgroup of the $SO(6)/SO(5)$ coset. In unitary gauge the Goldstone bosons eaten by the $W$ and $Z$ bosons ($\pi_1, \pi_2, \pi_3$) and the Goldstone boson eaten by $U(1)_\eta$ ($\pi_5$) are removed from the spectrum.

The SM quarks are introduced as chiral states external to the strong sector and are neutral under $SU(3)_{\tilde{c}} \times U(1)_\eta$. The left-handed (LH) top and bottom doublet can be embedded in a **6** of $SO(6)$, while the RH top quark is taken as a singlet. The specific embeddings are discussed in Sec. II A 2. The twin quarks required for the softening of the Higgs potential are also introduced as external chiral states, and are neutral under the SM gauge symmetries but triplets under $SU(3)_{\tilde{c}}$. The LH twin top is embedded in a **6** of $SO(6)$ such that it is charged under $U(1)_\eta$, while the RH twin top can be taken as a singlet under $SO(6)$ and thus has no $U(1)_\eta$ charge. It has been shown in [23,24] that with a $Z_2$ exchange symmetry fixing the Yukawa couplings of the Higgs with the top and twin top to be equal, the leading order contribution from the top quark to the Higgs potential is canceled. The $Z_2$ symmetry should generate Yukawa couplings that schematically look like

$$\mathcal{L} \supset \frac{f_\pi}{\sqrt{2}} y_t (s_h \bar{t}_L t_R + c_h \bar{\tilde{t}}_L \tilde{t}_R) \qquad (2.2)$$

where the top quarks are denoted by $t_{L,R}$, the twin tops by $\tilde{t}_{L,R}$, and $s_h$ and $c_h$ equal $\sin h/f_\pi$ and $\cos h/f_\pi$, respectively. This Lagrangian is invariant under $s_h \leftrightarrow c_h$ and $t \leftrightarrow \tilde{t}$ simultaneously. The leading order contribution to the Higgs potential is then $V(h) \sim y_t (s_h^2 + c_h^2) f_\pi^4$, which is independent of the Higgs field. Without the twin top coupling the leading order contribution would not vanish. Note here that it is important that the twin top quark does not have a vectorlike mass in the absence of the Higgs VEV, since then its contribution to the Higgs potential would scale like $(1 + c_h)^2$, and then the cancellation of $s_h^2$ at leading order would not occur. This is ensured by the fact that the RH twin top is not charged under $U(1)_\eta$. In this section we will show how this type of model can be realized in a 5D holographic scenario.

### A. The holographic model

Our starting point for the holographic description is the 5D RS model with the metric

$$ds^2 = \left( \frac{R}{z} \right)^2 \eta_{MN} dx^M dx^N \qquad (2.3)$$

where $z \equiv x^5$ is the extra dimensional coordinate. The extra dimensional space is cut off by two three branes; one in the UV at $z_0 = R$ and the other in the IR at $z_1 = R' \sim 1/\text{TeV}$. The RS model is conjectured to be dual to a strongly coupled gauge theory in 4D, whose conformal invariance is broken at the scale dictated by the position of the IR brane. In modeling a composite Higgs scenario which renders the Higgs sector natural we expect this scale to be close to 1 TeV. This IR scale is also known as the Kaluza-Klein (KK) scale and the excited modes of fields living in the 5D bulk are known as KK modes. The masses of the lightest KK modes are $\sim M_{KK} = 1/R'$ with the exact mass being determined by the particles spin and bulk dynamics. Once the IR scale is fixed, the UV scale is then related to the number of colors in the dual strongly interacting gauge theory through

$$\log(\Omega) = \frac{16\pi^2}{N} \frac{1}{g^2}, \qquad (2.4)$$

where $\Omega = R'/R$ is the 5D volume, $N$ is the number of colors, and $g$ is the electroweak coupling. The quantity $\Omega$ will play an important role in the determination of the Higgs mass and its decay constant later in the paper. Due to the 5D NDA condition for calculability the allowed values of $N$ are constrained to lay in $1 \ll N \lesssim 10$ [3,4,6], and in the work presented here we will allow $N$ in the range 5 to 10.

#### 1. The gauge sector

Details on the treatment of 5D gauge fields, including their boundary conditions on the branes, are given in Appendix B 1. We now use $\mathcal{G}$ to label the bulk gauge symmetry in the 5D model, while $\mathcal{H}$ and $\mathcal{H}_G$ label the gauge symmetries preserved on the IR and UV branes, respectively. The generators in $\mathcal{H}$ are those left unbroken by the spontaneous symmetry breaking induced by the strong sector, while the generators in $\mathcal{H}_G$ are those which are gauged in the effective theory. The Goldstone bosons in holographic models arise from the $A_5(x, z)$ components of the bulk gauge fields. We assume the following symmetry structure,

$$\mathcal{G} = SU(7) \times SO(6) \times U(1)_X$$
$$\mathcal{H} = SU(7) \times SO(5) \times U(1)_X$$
$$\mathcal{H}_G = SU(3)_{\tilde{c}} \times SU(3)_c \times SU(2)_L \times U(1)_Y \times U(1)_\eta,$$
$$\qquad (2.5)$$

which is similar to the gauge structure used for the holographic twin Higgs model in [36]. In $SU(7)$ there is the subgroup $SU(3)_c \times SU(3)_{\tilde{c}} \times U(1)_7 \times U(1)_{\tilde{7}}$, while in $SO(5)$ we have $SU(2)_L \times SU(2)_R$. The UV boundary conditions are chosen such that the hypercharge generator is given by the linear combination

$$Y = T_R^3 + X - \frac{4}{3} T^{\tilde{7}} \qquad (2.6)$$

with $T_3^R$ being the diagonal generator of $SU(2)_R$ and $T^{\tilde{7}}$ being the generator of $U(1)_{\tilde{7}} \subset SU(7)$. We denote the Abelian group formed from $T_R^3$ as $U(1)_R$. The gauging of $U(1)_\eta$ via UV boundary conditions ensures that the real singlet $\eta$ is eaten to form the longitudinal component of the massive spin-1 mode. Therefore the only massless scalars (pre-EWSB) here are the four Higgs degrees of freedom ar from the $SO(6) \to SO(5)$ breaking boundary conditions on the IR brane. To find the holographic action for these fields we start by writing down the most general $SU(7) \times SO(6) \times U(1)_X$ effective action to quadratic order for the spin-1 sector,

$$\mathcal{L} = \frac{P_T^{\mu\nu}}{2} [\Pi_0(p^2) \mathrm{Tr}(A_\mu A_\nu) + \Pi_1(p^2) \Sigma^T A_\mu A_\nu \Sigma \\ + \Pi_0^X(p^2) A_\mu^X A_\nu^X + \Pi_0^c(p^2) \mathrm{Tr}(A_\mu^c A_\nu^c)] \qquad (2.7)$$

where $A_\mu$ are the $SO(6)$ gauge fields and $A_\mu^{X,c}$ are the $U(1)_X$ and $SU(7)$ gauge fields, respectively. The Goldstone boson multiplet containing the Higgs field is defined in Eq. (2.1), which in unitary gauge is simply $\Sigma = (0, 0, 0, s_h, 0, c_h)^T$ with $s_h = \sin h/f_\pi$ and $c_h = \cos h/f_\pi$. The task is then to calculate these form factors from a 5D theory. In Appendix B 1 the form factors $\Pi_\pm$ have been defined with the $\pm$ referring to the IR boundary conditions. The form factors in the model under consideration can be written in terms of $\Pi_\pm$. Matching the action in Eq. (2.7), in the limit of $s_{\langle h \rangle} \to 0$, to the holographic effective action in the Appendix B we find

$$\Pi_0 = \frac{\Pi_+}{g_5^2} \qquad \Pi_1 = \frac{\Pi_- - \Pi_+}{g_5^2}$$

$$\Pi_0^X = \frac{\Pi_+}{g_{5,X}^2} \qquad \Pi_0^c = \frac{\Pi_+}{g_{5,c}^2} \qquad (2.8)$$

where $g_{5,c}$, $g_5$, and $g_{5,X}$ are the 5D gauge couplings of the $SU(7)$, $SO(6)$, and $U(1)_X$ gauge groups. These are the only form factors we need to describe the gauge fields and their interactions with the Higgs field at quadratic order. Expanding the action in Eq. (2.7), keeping only the gauged generators, we have

$$\mathcal{L} = \frac{P_T^{\mu\nu}}{2} \Big\{ W_\mu^+ \Big( \Pi_0 + \frac{s_h^2}{4} \Pi_1 \Big) W_\nu^- + Z_\mu \Big( \Pi_0 + \frac{s_h^2}{4 c_W^2} \Pi_1 \Big) Z_\nu \\ + A_\mu \Pi_0 A_\nu + B_\mu \Big( \Pi_0 + \frac{c_h^2}{4} \Pi_1 \Big) B_\nu \\ + A_\mu^c \Pi_0^c A_\nu^c + A_\mu^{\tilde{c}} \Pi_0^{\tilde{c}} A_\nu^{\tilde{c}} \Big\}. \qquad (2.9)$$

The $U(1)_\eta$ boson is denoted by $B_\mu$, the photon by $A_\mu$, and the $SU(3)_c$ and $SU(3)_{\tilde{c}}$ gauge bosons by $A_\mu^c$ and $A_\mu^{\tilde{c}}$, respectively. The $s_W$ and $c_W$ symbols denote sine and cosine functions of the Weinberg angle.

At low energies the $\Pi_\pm$ form factors behave as

$$\Pi_+(p_E^2 \sim 0) \simeq -p_E^2 R \log \Omega, \qquad \Pi_-(p_E^2 \sim 0) \simeq -\frac{2R}{R'^2}. \qquad (2.10)$$

Requiring the proper normalization of Eq. (2.9) implies that $g_5^2 = g^2 R \log \Omega$, $g_5^2 = g'^2 R \log \Omega$, and $g_{5,c}^2 = g_c^2 R \log \Omega$, with $g$, $g'$, and $g_c$ being the SM gauge couplings. The SM gauge fields couple to the Higgs with a term $\sim s_h$ while the $U(1)_\eta$ field couples with a term $\sim c_h$, indicating that the electroweak gauge group is unbroken in the $\langle h \rangle = 0$ limit while the $U(1)_\eta$ gauge symmetry remains broken. The decay constant of the Higgs field is identified from the low energy limit of $\Pi_1$,

$$\Pi_1(p_E = 0) = -\frac{f_\pi^2}{2} \Rightarrow f_\pi^2 = \frac{4}{g^2 R'^2 \log \Omega} = \frac{N}{4\pi^2 R'^2}. \qquad (2.11)$$

This implies that the SM Higgs VEV ($v$) is related to $\sin \frac{\langle h \rangle}{f_\pi} \equiv s_{\langle h \rangle}$ via $s_{\langle h \rangle} = \frac{v}{f_\pi}$, and that the mass of the $U(1)_\eta$ gauge boson is

$$m_{B'} = \frac{g}{4} \sqrt{f_\pi^2 - v^2}. \qquad (2.12)$$

The couplings between the Higgs and the electroweak gauge bosons have the usual corrections one encounters in composite Higgs models, $g_{hVV} = g_{hVV_{\mathrm{SM}}} \sqrt{1 - s_{\langle h \rangle}^2}$. The $U(1)_\eta$ gauge coupling is the same as the electroweak gauge coupling and the coupling between the Higgs and the $U(1)_\eta$ boson is given by $g_{hBB} = -g_{hWW_{\mathrm{SM}}} \sqrt{1 - s_{\langle h \rangle}^2}$. The gauge coupling for $U(1)_\eta$ is not required to be the same as the electroweak gauge coupling, as this could be altered in the 5D theory by adding UV brane kinetic terms for the gauge fields.

### 2. The top sector

We embed the LH quark doublet and the LH twin top in a $(7, 6)_{\frac{2}{3}}$ of $SU(7) \times SO(6) \times U(1)_X$, while the RH top and RH twin top are embedded as $SO(6)$ singlets in a $(7, 1)_{\frac{2}{3}}$. The $SU(7)$ subgroups, $SU(3)_c \times SU(3)_{\tilde{c}} \times U(1)_7 \times U(1)_{\tilde{7}}$, are labeled such that the twin quarks are charged under $SU(3)_{\tilde{c}} \times U(1)_{\tilde{7}}$, and the SM quarks are charged under $SU(3)_c \times U(1)_7$. Specifically, the SM quarks are in the $(3, 1)_{\frac{1}{3}, 0}$ representation, and the twin top quarks are in the $(\mathbf{1}, \mathbf{3})_{0, \frac{1}{2}}$ representation. The $SO(6)$ vectors containing the LH top quark ($t_L$) and its twin ($\tilde{t}_L$) are

$$q_L = \frac{1}{\sqrt{2}} \begin{pmatrix} ib_L \\ b_L \\ it_L \\ -t_L \\ 0 \\ 0 \end{pmatrix}, \qquad \tilde{t}_L = \frac{1}{\sqrt{2}} \begin{pmatrix} 0 \\ 0 \\ 0 \\ 0 \\ i\tilde{t}_L \\ \tilde{t}_L \end{pmatrix}. \qquad (2.13)$$

The first four components of these vectors are $SO(4)$ multiplets and thus carry $SU(2)_L$ charge, while the last two components are vectors of $SO(2)_\eta$, i.e., charged under $U(1)_\eta$. Under the $SU(2)_L \times U(1)_R \times U(1)_\eta$ subgroup of $SO(6)$, with $U(1)_R$ being the diagonal subgroup of $SU(2)_R$, the above embeddings imply that the SM doublet has charges $\mathbf{2}_{-\frac{1}{2},0}$, and the twin top has charges $\mathbf{1}_{0,\frac{1}{2}}$. To summarize, the subgroups of the bulk gauge symmetry relevant for the quark sector are

$$\mathcal{H}_F = SU(3)_{\tilde{c}} \times SU(3)_c \times SU(2)_L \times U(1)_R \times U(1)_X$$
$$\times U(1)_{\tilde{\eta}} \times U(1)_\eta \qquad (2.14)$$

under which the SM quarks and twin quarks have charges

$$(t_L, b_L): (\mathbf{1}, \mathbf{3}, \mathbf{2})_{-\frac{1}{2}, \frac{2}{3}, 0, 0}$$
$$t_R: (\mathbf{1}, \mathbf{3}, \mathbf{1})_{0, \frac{2}{3}, 0, 0}$$
$$\tilde{t}_L: (\mathbf{3}, \mathbf{1}, \mathbf{1})_{0, \frac{2}{3}, \frac{1}{2}, \frac{1}{2}}$$
$$\tilde{t}_R: (\mathbf{3}, \mathbf{1}, \mathbf{1})_{0, \frac{2}{3}, \frac{1}{2}, 0}. \qquad (2.15)$$

With $Y = T_R^3 + X - \frac{4}{3}T^{\tilde{7}}$ from Eq. (2.6), these embeddings will result in the SM quarks having their appropriate hypercharges while the twin quarks will be neutral under SM gauge symmetries. This is the motivation behind choosing hypercharge as the linear combination in Eq. (2.6). To properly include the bottom quark in this setup we are required to introduce a second electroweak doublet in an $SO(6)$ vector as in Eq. (2.13) but with a $U(1)_X$ charge of $-1/3$. This is required in order for the bottom quark to obtain a mass via EWSB, see Appendix A.1 of [6] for more details. Therefore, the second quark doublet is embedded in a $(\mathbf{7}, \mathbf{6})_{-\frac{1}{3}}$ of $SU(7) \times SO(6) \times U(1)_X$, while the RH bottom is embedded as an $SO(6)$ singlet in a $(\mathbf{7}, \mathbf{1})_{-\frac{1}{3}}$. Note that we do not introduce a chiral twin bottom quark here, since due to its charge under $U(1)_X$ it would carry hypercharge and would result in serious phenomenological issues with the model. The effects of the $Z_2$ breaking from the bottom quark contributions to the Higgs potential are negligible in comparison to those from electroweak gauge bosons and the top quark, and hence the effects of the bottom quark do not require suppression from the twin mechanism.

In the 5D holographic model the external chiral fermions arise as massless modes of a bulk 5D Dirac fermion. Thus in the effective theory each chiral fermion is accompanied by a tower of vector-like fermions with the same charges under the bulk gauge symmetries. So the SM quarks will be accompanied by a tower of colored vectorlike states while the twin quarks will be accompanied by a tower of uncolored vectorlike states. The $Z_2$ symmetry between the quarks and their twin partners is enforced by embedding them in a single representation of the 5D gauge symmetry. This means that there will also be heavy KK states charged under both twin and SM gauge symmetries. To summarize the 5D quark content, we include the third generation quarks in 5D fermion multiplets with the following charge assignments under the bulk $SU(7) \times SO(6) \times U(1)_X$ gauge symmetry,

$$\xi_q = (\mathbf{7}, \mathbf{6})_{\frac{2}{3}} \qquad \xi_{q'} = (\mathbf{7}, \mathbf{6})_{-\frac{1}{3}}$$
$$\xi_u = (\mathbf{7}, \mathbf{1})_{\frac{2}{3}} \qquad \xi_d = (\mathbf{7}, \mathbf{1})_{-\frac{1}{3}}. \qquad (2.16)$$

As explained in [6], the additional massless states ar from the doubling of the LH SM quarks is cured by introducing a mass mixing on the UV brane lifting the mass of one linear combination of the bulk doublet fields. Due to the fact that the top quark Yukawa coupling is significantly larger than that of the bottom quark, we can safely neglect the effects of the bottom sector when we study the Higgs potential and top-partner spectra. Therefore from this point on we will only consider the $\xi_q$ and $\xi_u$ multiplets in our study. The complete 5D $SO(6)$ multiplets for the top quark and the twin top quark can be written as,

$$\xi_q^c = \frac{1}{\sqrt{2}} \begin{pmatrix} iB - iX_{5/3} \\ B + X_{5/3} \\ iT + iX_{2/3} \\ -T + X_{2/3} \\ iT_+ - iT_- \\ T_+ + T_- \end{pmatrix} \qquad \xi_q^{\tilde{c}} = \frac{1}{\sqrt{2}} \begin{pmatrix} i\tilde{B} - i\tilde{X}_{5/3} \\ \tilde{B} + \tilde{X}_{5/3} \\ i\tilde{T} + i\tilde{X}_{2/3} \\ -\tilde{T} + \tilde{X}_{2/3} \\ i\tilde{T}_+ - i\tilde{T}_- \\ \tilde{T}_+ + \tilde{T}_- \end{pmatrix}.$$
$$(2.17)$$

The subscripts on the $X$ and $\tilde{X}$ fields label the EM charge, whereas the $T_\pm$ and $\tilde{T}_\pm$ fields have $2/3$ $U(1)_X$ charges with the $\pm$ referring to $\pm \frac{1}{2}$ $U(1)_\eta$ charges. The different states in these multiplets have both LH and RH components, and we will use projection operators ($P_{L,R}$) to identify between different chiralities. We identify the LH top quark with the zero mode of $P_L T$, and the LH twin top with the zero mode of $P_L \tilde{T}_+$. By choice we enforce that all other components of these multiplets do not have massless modes, this is done through the introduction of Lagrange multiplier fields on the UV brane which couple linearly to these components [38], and is analogous to the Dirichlet UV boundary conditions for components of the multiplets without zero modes. For the RH chiral quarks embedded as singlets of $SO(6)$ the boundary conditions are chosen so that the chiral zero mode is RH. To summarize, under $SU(3)_c \times SU(3)_{\tilde{c}} \times SU(2)_L \times U(1)_Y$ the top and twin top fields have the following charges,

$$
\begin{aligned}
t_L &= (\mathbf{3},\mathbf{1},\mathbf{2})_{1/6}, & \tilde{t}_L &= (\mathbf{1},\mathbf{3},\mathbf{1})_0, \\
t_R &= (\mathbf{3},\mathbf{1},\mathbf{1})_{2/3}, & \tilde{t}_R &= (\mathbf{1},\mathbf{3},\mathbf{1})_0.
\end{aligned}
\tag{2.18}
$$

Although we have used the $U(1)_\eta$ charges to identify some of the quarks in the $SO(6)$ multiplets, this is not a particularly useful labeling in the effective theory where $U(1)_\eta$ is spontaneously broken. When describing the mass spectra of the composite resonances it is better instead to use $T_{P,M} = \frac{1}{\sqrt{2}}(T_+ \pm T_-)$ and $\tilde{T}_{P,M} = \frac{1}{\sqrt{2}}(\tilde{T}_+ \pm \tilde{T}_-)$. At the zero mode level the LH components of $\tilde{T}_{P,M}$ reduce to $P_L \tilde{T}_P = P_L \tilde{T}_M = \frac{1}{\sqrt{2}}\tilde{t}_L$.

In Appendix B 2 we give a brief discussion of how the holographic technique is applied to 5D fermion fields. In the 5D model the bulk gauge symmetry is broken to $SO(5)$ on the IR brane, therefore we will allow mass mixings ($\tilde{m}$) between the multiplets to exist on the IR brane which respect only the $SU(7) \times SO(5) \times U(1)_X$ invariance. Analogously to Eq. (B6), the 5D Lagrangian for this scenario can be written as

$$
\begin{aligned}
\mathcal{L} = \int dx^5 \sqrt{|g|} \Big\{ &\frac{i}{2} \bar{\xi}_q \gamma^M \partial_M \xi_q - \frac{i}{2}(\partial_M \bar{\xi}_q)\gamma^M \xi_q - m_q \bar{\xi}_q \xi_q \\
&\times \frac{i}{2} \bar{\xi}_u \gamma^M \partial_M \xi_u - \frac{i}{2}(\partial_M \bar{\xi}_u)\gamma^M \xi_u - m_u \bar{\xi}_u \xi_u - \delta(z-R')\tilde{m}(\bar{\xi}_{q1L}\Sigma_0\xi_{uR} + \bar{\xi}_{uR}\Sigma_0^T\xi_{q1L}) \\
&\times \delta(z-R)\frac{1}{2}(\bar{\xi}_{q1L}\xi_{q1R} - \bar{\xi}_{uL}\xi_{uR}) - \delta(z-R')\frac{1}{2}(\bar{\xi}_{q1L}\xi_{q1R} - \bar{\xi}_{uL}\xi_{uR}) \Big\}
\end{aligned}
\tag{2.19}
$$

where $\Sigma_0^T = (0,0,0,0,0,1)$. The 5D masses $m_q$ and $m_u$ can be written in terms of dimensionless quantities $m_{q,u} = c_{q,u}/R$. The UV boundary conditions for the 5D fields choose a LH source field for $\xi_q$, and a RH source field for $\xi_u$,

$$
\xi_q(p,R) \equiv \xi_{q1L}(p), \qquad \xi_u(p,R) \equiv \xi_{uR}(p).
\tag{2.20}
$$

The IR boundary conditions are derived from the mass mixing between the multiplets mediated by $\tilde{m}$. These mass mixings are analogous to the partial compositeness mixing commonly used in 4D implementations of composite Higgs models. The goal now is to use this 5D model to calculate the effective action for the Yukawa sector. The most general effective action for the $\xi_{q,u}$ multiplets at quadratic order is

$$
\begin{aligned}
\mathcal{L} = &\bar{\xi}_q \slashed{p}(\Pi_0^q(p) + \Pi_1^q(p)\Sigma\Sigma^T)\xi_q + \bar{\xi}_u \slashed{p}\Pi_0^t(p)\xi_u \\
&+ \bar{\xi}_q M_1^t(p)\Sigma\xi_u + \text{H.c.}
\end{aligned}
\tag{2.21}
$$

Utilizing the holographic techniques outlined in Appendix B 2 for the model described here, we can match the holographic action in Eq. (B8) to the action in Eq. (2.21) in the limit that $\Sigma = \Sigma_0$. In doing so we find

$$
\begin{aligned}
\Pi_0^q &= \Pi_1^f(\tilde{m}=0) & \Pi_1^q &= (\Pi_1^f - \Pi_1^f(\tilde{m}=0)) \\
\Pi_0^t &= \Pi_2^f & M_1^t &= M^f
\end{aligned}
\tag{2.22}
$$

with the 5D form factors $\Pi_{1,2}^f$ and $M^f$ being given in the Appendix B. Expanding the action in Eq. (2.21) with $\langle h \rangle \neq 0$, keeping only the top and twin top, we have

$$
\begin{aligned}
\mathcal{L} = &\bar{t}_L \slashed{p}\Big(\Pi_0^q + \frac{1}{2}\Pi_1^q s_h^2\Big)t_L + \bar{\tilde{t}}_L \slashed{p}\Big(\Pi_0^q + \frac{1}{2}\Pi_1^q c_h^2\Big)\tilde{t}_L \\
&+ \bar{t}_R \slashed{p}\Pi_0^t t_R + \bar{\tilde{t}}_R \slashed{p}\Pi_0^t \tilde{t}_R - \frac{M_1^t}{\sqrt{2}}(\bar{t}_L t_R s_h + \bar{\tilde{t}}_L \tilde{t}_R c_h) + \text{H.c.}
\end{aligned}
\tag{2.23}
$$

with the form factors now determined by (2.22) with the explicit expressions being given in the Appendix B 2. The $Z_2$ ($s_h \leftrightarrow c_h$, $t_L \leftrightarrow \tilde{t}_L$) symmetry is now explicit in the top Yukawa couplings, the implications of which we will study in the next chapter. Other crucial features that we should extract from the effective action are the masses of the lightest vectorlike top-partners, i.e., $T_{(L,R)}$, $X_{(L,R)}$, $T_{M(L,R)}$ and $T_{P(L,R)}$ and their twin counterparts. In the limit of $\langle h \rangle = 0$ these masses are given by

$$
\begin{aligned}
m_T &= \text{zeros}(\Pi_0^q) \\
m_{X_{2/3}} &= m_{T_M} = \text{poles}(\Pi_0^q) \\
m_{T_P} &= \text{poles}\Big(\Pi_0^q + \frac{1}{2}\Pi_1^q\Big)
\end{aligned}
\tag{2.24}
$$

where the $\langle h \rangle \neq 0$ effects are small for top-partners at the TeV scale. Note that apart from the chiral modes, the top and twin top sectors have the same spectra in the $\langle h \rangle \to 0$ limit. The top and twin top masses are given by

$$m_t = \frac{1}{\sqrt{2}} \frac{M_1^t s_{\langle h \rangle}}{\sqrt{\Pi_0^q} \sqrt{\Pi_0^q + \frac{1}{2} \Pi_1^q s_{\langle h \rangle}^2}} \Bigg|_{p^2 \simeq 0}$$

$$m_{\tilde{t}} = m_t |_{s_{\langle h \rangle} \to c_{\langle h \rangle}} = m_t \frac{\sqrt{f_\pi^2 - v^2}}{v}. \qquad (2.25)$$

These masses are strongly sensitive to the 5D mass parameters $c_{q,u}$. In a KK decomposition the localization of the fermion zero modes actually depends exponentially on these parameters, with $c_q > 0$ indicating that the SM doublet zero mode is localized away from the IR region of the extra dimension, with the opposite being true for $c_u$ and the singlet zero mode. We can derive the coupling of the twin top quark to the Higgs from the above relation, finding $y_{\tilde{t}} \simeq -s_{\langle h \rangle} y_t$.

From the perspective of the effective theory we have a global $SO(6)$ symmetry breaking to $SO(5)$, and a gauge symmetry $SU(3)_c \times SU(3)_{\tilde{c}} \times SU(2)_L \times U(1)_Y \times U(1)_\eta$ breaking to $SU(3)_c \times SU(3)_{\tilde{c}} \times U(1)_Q$. An issue quickly arises when we look at the chiral fermion content of the theory, as since the LH twin top is the only field charged under the $U(1)_\eta$ gauge symmetry chiral anomalies arise in the twin sector through triangle diagrams involving the $U(1)_\eta$ field on one or all three of the external legs. A simple way to avoid this is to assume that for the LH charm quark, for example, there is a twin quark with charge $-1/2$ under $U(1)_\eta$. This can occur naturally through the $SO(6)$ vector representation and the form of the contribution of the twin quark to the Higgs potential is independent of the charge under $U(1)_\eta$, although this would not be an issue anyway due to the smallness of the charm quark contributions.

However when constructing models of EWSB in interval or orbifolded spacetimes it is possible that, although the theory is anomalyfree from the perspective of the effective theory, there are brane-localized anomalies ar due to a incomplete or inconsistent description of the model. Much work has been done in the study of anomalous symmetries in 5D constructions [39,40], and in [41] the anomalous structure of composite Higgs models is studied. A simple example of a brane-localized chiral gauge anomaly arises when considering a bulk $U(1)$ gauge symmetry with a single bulk fermion $\Psi(x, z)$ charged under the bulk gauge symmetry with charge $Q$. We assume for now that the gauge symmetry is unbroken on the branes and thus the spectrum contains a massless $U(1)$ gauge field. For the fermion we consider two choices of boundary conditions on each brane, Neumann or Dirichlet, for the LH component, i.e., $\Psi_R|_{\text{brane}} = 0$ or $\Psi_L|_{\text{brane}} = 0$, since the RH component must have a boundary condition opposite to that of the LH component. The 5D Lorentz symmetry imposes chirality in the 5D bulk and therefore the only source of chiral breaking is on the branes, for this reason any gauge anomalies must be restricted to the branes. On variation of the effective action this model results in brane-localized chiral anomalies of the form $\sim \int d^4x dz \Lambda(x, z) \frac{Q^3}{96\pi^2} F_{\mu\nu} \tilde{F}^{\mu\nu} \times \delta(z \pm z_i)$ where the coefficient $= \pm 1$ for Neumann/ Dirichlet boundary conditions at $z_i = R, R'$. With Neumann boundary conditions on each brane we have a massless LH fermion in the spectrum. In this case the brane-localized anomalies match what one would expect if only the effective theory containing the massless LH fermion and the $U(1)$ gauge field was considered. The same situation arises if we choose Dirichlet boundary conditions on both branes, with the only difference being that we now have a RH massless fermion rather than LH. An interesting case arises when we have mixed boundary conditions, i.e., Neumann on one brane and Dirichlet on the other. Here there is no massless fermion in the spectrum, and the anomalies cancel upon integrating over the whole extra dimension. However they are still nonzero in the full theory, i.e., on the branes, and therefore gauge invariance is broken explicitly and the theory is inconsistent.

Various solutions are described in [39], such as modifying the boundary conditions of the gauge fields for which an anomaly is present, introducing additional bulk or brane fermions, or introducing a bulk Chern-Simons term. The same analysis can be extended to the non-Abelian case in a straightforward way, where to obtain a consistent 5D theory one must again show that the boundary conditions for the fermions and gauge fields on each brane are anomalyfree. In the model described in this paper we have a bulk gauge symmetry $SU(7) \times SO(6) \times U(1)_X$. On the IR brane this gauge symmetry is broken via boundary conditions to $SU(7) \times SO(5) \times U(1)_X$, with the boundary conditions of the bulk fermions also respecting the same global symmetry. On the UV brane the bulk gauge symmetry is then broken to the gauge symmetry of the effective theory. Without the introduction of additional fermionic content, either through the introduction of the full spectrum of SM quarks and leptons in the bulk or even some additional BSM states, the model described here does give rise to anomalies on the branes. This is to be expected since even in the SM anomaly cancellation requires contributions from both quarks and leptons. The only chiral fermions in the theory are the SM fermions plus the LH and RH twin top quark, and possibly other chiral twin fermions. We find that if the chiral anomalies cancel when only the chiral states in the effective theory are considered, any additional brane-localized gauge anomalies ar from bulk fermions with mixed boundary conditions can be canceled by introducing additional bulk fermions also with mixed boundary conditions.[1] It is important to note that this is no way changes the makeup of low energy theory, since the lightest states in the spectrum arising from a bulk fermion with mixed boundary conditions have a mass at the KK scale. The contributions of these additional KK states to the Higgs potential can also be assumed to be negligible since large contributions only arise through a large explicit breaking of the global symmetry, which is in no way necessary of these states. Thus the problem reduces to finding a model in which the anomalies in the effective theory cancel. This can be achieved, as pointed out previously, through the introduction of an additional LH twin quark charged under twin QCD and $U(1)_\eta$, but with a $U(1)_\eta$ charge opposite to that of the twin top quark. For example, a second generation a twin quarks with the twin LH charm quark having a $U(1)_\eta$ charge opposite to that of the twin LH top quark, and a twin RH charm quark embedded in the same way as the twin RH top quark. These states are irrelevant for the calculation of the Higgs potential in the next section.

---

[1]This is assuming that the brane-localized anomalies are due to bulk fermions, as examples where anomalies arise due to brane-localized fermions (as in the $SU(2)$ example of [39]) cannot generally be cured in this way.

## III. THE HIGGS POTENTIAL

Given the effective actions that we have derived for the gauge and fermion fields, the one-loop Coleman-Weinberg potential for the Higgs field can be written as

$$V(h) = V_G(h) + V_{t,\tilde{t}}(h)$$

$$V_G(h) = \frac{3}{2} \int \frac{d^4 p_E}{(2\pi)^4} \left\{ 2\log\left[1 + \frac{s_h^2}{4}\frac{\Pi_1}{\Pi_0}\right] + \log\left[1 + \frac{s_h^2}{4c_W^2}\frac{\Pi_1}{\Pi_0}\right] + \log\left[1 + \frac{c_h^2}{4}\frac{\Pi_1}{\Pi_0}\right] \right\}$$

$$V_{t,\tilde{t}}(h) = -2N_c \int \frac{d^4 p_E}{(2\pi)^4} \left\{ \log\left[1 + \frac{s_h^2}{2}\frac{\Pi_q^1}{\Pi_q^0}\right] + \log\left[1 + \frac{s_h^2}{2}\frac{(M_t^1)^2}{\Pi_t^0(\Pi_q^0 + \frac{s_h^2}{2}\Pi_q^1)}\right] \right\} + (s_h \to c_h).$$
(3.1)

We see here that there is an exchange symmetry between $s_h \leftrightarrow c_h$ in the top quark contribution but not in the contribution from the gauge bosons. This is an explicit breaking of the $Z_2$ symmetry that is present in the model by construction, and is crucial for obtaining a realistic model of EWSB. Indeed this is one major advantage of this model over other models, such as those based on an $SO(8)/SO(7)$ coset, where additional sources of $Z_2$ breaking are introduced by hand. Because the bottom quark does not have a twin partner means that there will also be violations of the $Z_2$ symmetry for the down quarks, however these effects are negligible in comparison to those in the electroweak sector. In studying potentials of this type one option is to expand the logarithms so that we have a polynomial in $s_h^2$. The leading term in the prefactor of the $s_h^2$ term in $V_{t,\tilde{t}}$ would then be $\sim (M_t^1)^2$, however due to our $c_h^2$ contribution from the twin top this term vanishes and the leading contributions are $\sim (M_t^1)^4$. A spurious IR divergence enters in this term solely due to the expansion of the logarithm [12]. When the $(M_t^1)^2$ term is present the problem is avoided by introducing an IR cutoff, to which the results are not sensitive. However when this term is not present and the leading contribution is the $(M_t^1)^4$ term, it will be beneficial to use another method for evaluating the Higgs potential which does not introduce IR divergences. To do this we will integrate the whole potential numerically while scanning over values of $s_{\langle h \rangle}$ to find the minimum of the potential and the Higgs mass.

To begin the analysis we will simply look at the potential as a function of $s_h$. We will fix $R' = 1/M_{KK} = 1/(1500 \text{ GeV})$, $N = 8$, and $c_q = 0.25$. The mass parameter $\tilde{m}$ is fixed such that $m_t = \frac{1}{\sqrt{2}} v$ and $c_u$ is varied in the range $[-0.4, 0.4]$. From Fig. 1 we can see that for values of $c_u$ closer to $-0.4$ the Higgs mass and vacuum expectation value increase, as does the value of $\tilde{m}$ required to achieve $m_t = \frac{1}{\sqrt{2}} v$. For $M_{KK} = 1500 \text{ GeV}$ the lightest spin-1 resonances are at a mass $\simeq 3.6$ TeV, whereas the masses of the lightest fermionic states depend on the 5D fermion mass parameters and will be the study of the next section. It is noteworthy that there is no tuning required to obtain a light Higgs mass and a small vacuum expectation value, as can be seen through the varying of $c_u$. This is one of the striking results of the $Z_2$ symmetry present in this model, and in twin Higgs models in general. In fact one often finds that the Higgs mass is too light in comparison to the vacuum expectation value leading to the requirement of additional $Z_2$ breaking terms. Again, one of the benefits of the model studied in this paper is that the $Z_2$ breaking is already by construction, since the discrete symmetry is present only in the top sector. This is due to the model producing a value of the quartic coupling which is too small, and is a general feature of twin Higgs models. Work has been done in building models in which the Higgs has a quartic interaction at tree-level, but a mass only at loop level [42].

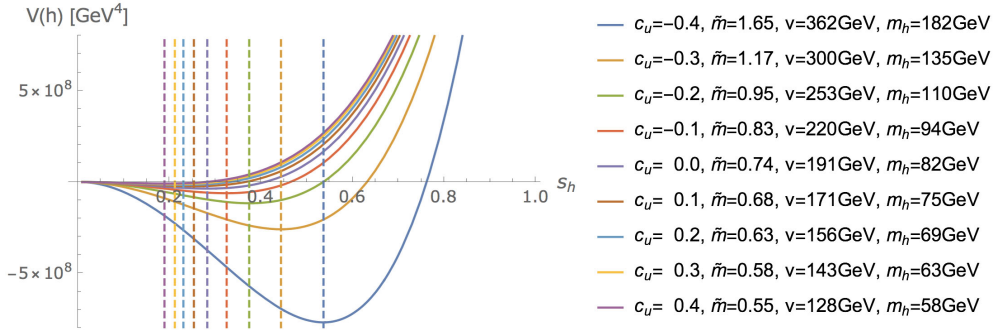

FIG. 1. The Higgs potential is plotted as a function of $s_h$ for $c_q = 0.25$, $N = 8$, and $M_{KK} = 1500$ GeV. The parameter $\tilde{m}$ is chosen such that $m_t = \frac{1}{\sqrt{2}} v$, and $c_u$ is varied in the range $[-0.4, 0.4]$. The vertical lines indicate the minimum of each curve, and in the legend we have included the Higgs mass and vacuum expectation value for each potential.

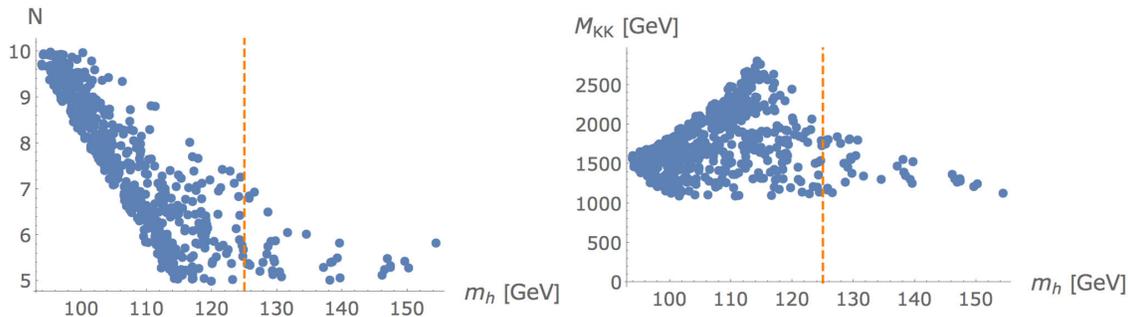

FIG. 2. In these plots we show how the Higgs mass depends on the number of colors $N$ and $M_{KK}$. We have scanned over $-0.45 \leq c_u \leq 0.45$, $0.15 \leq c_q \leq 0.4$, $1100 \text{ GeV} \leq M_{KK} \leq 4000 \text{ GeV}$, $5 \leq N \leq 10$. And the top mass and Higgs vacuum expectation value are fixed to their known values.

## A. Numerical scan

The free parameters in the model are

$$M_{KK}, \quad c_q, \quad c_u, \quad \tilde{m}, \quad N. \quad (3.2)$$

In this section we present the results of a scan over the parameter space, where the brane mass parameter $\tilde{m}$ is fixed to reproduce the top quark mass. The parameter ranges that we have scanned over are $-0.45 \leq c_u \leq 0.45$, $0.15 \leq c_q \leq 0.4$, $1100 \text{ GeV} \leq M_{KK} \leq 4000 \text{ GeV}$, $5 \leq N \leq 10$. The results are summarized in Figs. 2 and 3. In Fig. 2 we show how the Higgs mass depends on the number of colors $N$ and on $M_{KK}$. Interestingly we find that in order to reproduce the correct Higgs mass we require the number of colors to be less than approximately 7. This is due to the fact that $m_h$ scales inversely with $f_\pi$, and in models of neutral naturalness such as the twin Higgs, it is a general feature that Higgs mass is too small. In this scenario we see that this requires us to choose $N$ in a particular range, as opposed to introducing additional sources of $Z_2$ breaking terms as is done in other cases. We also see that in general a Higgs mass of 125 GeV picks out points in parameter space where $M_{KK}$ is $\lesssim 2000$ GeV.

In Fig. 3 we have shown the top-partner mass spectra for each of the points in the scan. Interestingly we see that having a Higgs mass of 125 GeV does not require top-partners in the mass ranges excluded by recent LHC analyses, i.e., $\lesssim 1.4$ TeV. This happens because the twin top, at a mass of $m_t \cot\langle h \rangle / f_\pi$, cancels the leading order contributions to the Higgs potential. A large constraint on the parameter space comes from the current bound on the decay constant of the composite Higgs, $f_\pi \gtrsim 600$ GeV [43].

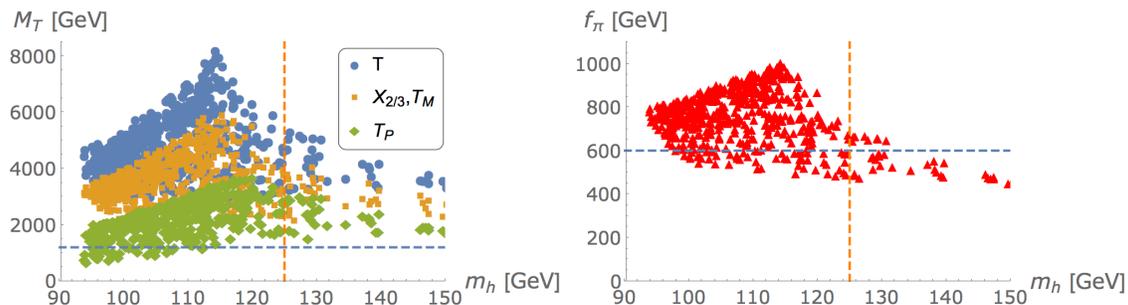

FIG. 3. In these plots we have show how the Higgs mass depends on the top-partner mass spectra and the decay constant. We have scanned over $-0.45 \leq c_u \leq 0.45$, $0.15 \leq c_q \leq 0.4$, $1100 \text{ GeV} \leq M_{KK} \leq 4000 \text{ GeV}$, $5 \leq N \leq 10$. And the top mass and vacuum expectation value are fixed to their known values. The horizontal blue dashed lines indicate the $M_T = 1200 \text{ GeV}$ and $f_\pi = 600$ GeV marks.

From the second plot in Fig. 3 we see that less than half of the viable points at $m_h \simeq 125$ GeV pass this constraint. The points which do pass this constraint are those with the larger values of $N$ and $M_{KK}$. The fine-tuning present in obtaining a realistic EWSB can be estimated as $\sim (v/f_\pi)^2$, and in our case this is in the range $\sim 12$–$17\%$.

## IV. CONCLUSIONS

In this paper we have presented a holographic description of a neutral natural composite Higgs model based on the $SO(6)/SO(5)$ coset. The model that we have studied is similar to those presented in [23,24], and the results we have derived through the holographic calculations agree well with those derived in these papers. We studied how the Higgs mass and top-partner spectra depend on the parameters in the model once the Higgs vacuum expectation value and the top quark mass are fixed. We found that the model can easily reproduce the SM observables without predicting colored top-partners lighter than $\sim 1500$ GeV, and without requiring additional sources of $Z_2$ breaking not already present in the model. However we do require a Higgs decay constant $\lesssim 700$ GeV, and therefore this scenario could be ruled out with more accurate measurements of the couplings of the Higgs boson to the electroweak gauge bosons. It is worth noting that in [23,24] no bound on $f_\pi$ was required, however similar features were observed in the holographic twin Higgs model presented in [36], where additional $Z_2$ breaking terms were introduced which increased the allowed range of $f_\pi$.

The $U(1)_\eta$ boson and the twin top only couple directly to the SM through the Higgs boson. The couplings are given in the main text, with the coupling of the $U(1)_\eta$ boson being similar to that of the SM electroweak bosons, and the Yukawa coupling of the twin top being suppressed with respect to the top Yukawa coupling by a factor of $s_{\langle h\rangle}$. Pair production of the twin sector states can only occur through an off-shell Higgs, and is thus very suppressed. These states can decay directly to other twin sector quarks, or partially to the SM via Higgs-strahlung. Because the twin states are not colored and only couple to the SM through the Higgs it will be difficult to probe these states at the LHC. Dark matter searches with monojet plus missing energy signals may be relevant for the phenomenology, as the states can be pair-produced from an off-shell Higgs in association with a jet and decay to twin sector particles which escape the detector. Future colliders such as the FCC [44] may be able to directly produce the colored top-partners with masses in the multi-TeV range, these could then be detected directly through their decay to SM states. However, some of these KK modes couple to both SM gluons and to the $U(1)_\eta$ boson, therefore processes involving intermediate KK quarks decaying to twin and SM states can occur. As mentioned in the main text, the most stringent experimental constraint on these models is on $v/f_\pi$, which is probed through the coupling of the Higgs to electroweak gauge bosons. Another interesting study, which we will pursue in another paper, is the phenomenology of the radion [45] and possibly the KK graviton [46] in this model and in holographic twin Higgs models in general. In neutral naturalness models the decays of the radion and KK graviton states to twin-sector particles could significantly change the phenomenological bounds, however we expect this to be more relevant for the radion than the KK graviton due to the radion having a much smaller mass than the lightest KK graviton state.

In conclusion, we have proposed a consistent holographic description of a neutral naturalness composite Higgs model based on an $SO(6)/SO(5)$ coset. We have shown that without the introduction of any additional sources of $Z_2$ breaking, commonly used in other twin Higgs models, the correct Higgs mass, Higgs vacuum expectation value, and top quark mass can be reproduced without the need for light colored top-partners. There is of course additional particle content associated with the neutral naturalness mechanism, but this is largely unimportant for LHC phenomenology. The main condition required is that $N \lesssim 7$, in which case $(v/f_\pi)^2$ in the range $\sim 12$–$17\%$. We then found that the lightest colored top-partners have masses $\gtrsim 1500$ GeV, above the bounds set by current LHC analyses.

## ACKNOWLEDGMENTS

The author acknowledges funding from UK Engineering and Physical Sciences Research Council (EPSRC) Grant No. EP/P005217/1, and would like to thank Stephan Huber and Aqeel Ahmed for useful comments on the draft. In the later stages of the project the author also acknowledges the financial support from the Slovenian Research Agency (research core funding No. P1-0035 and J1-8137).

## APPENDIX A: $SO(6)/SO(5)$ ALGEBRA

The generators of the $SO(6)$ algebra can be written as

$$T^{\hat{a}}_{ij} = -\frac{i}{\sqrt{2}}(\delta^{\hat{a}i}\delta^{6j} - \delta^{\hat{a}j}\delta^{6i})$$

$$T^{a}_{L,R,ij} = -\frac{i}{2}\left(\frac{1}{2}\epsilon^{abc}(\delta^{bi}\delta^{cj} - \delta^{bj}\delta^{ci}) \pm (\delta^{ai}\delta^{4j} - \delta^{aj}\delta^{4i})\right)$$

$$T^{\alpha}_{ij} = -\frac{i}{\sqrt{2}}(\delta^{\alpha i}\delta^{5j} - \delta^{\alpha j}\delta^{5i}) \qquad (A1)$$

where $a = 1, 2, 3$ labels the three generators for the $SU(2)_L$ and $SU(2)_R$ subgroups, $\hat{a} = 1, \ldots, 5$ labels the broken generators in the coset, and the remaining generators are given by $\alpha = 1, \ldots, 4$. The Higgs degrees of freedom are formed from the $\hat{a} = 1, \ldots, 4$ generators while the $U(1)_\eta$ symmetry is generated by $\hat{a} = 5$.

## APPENDIX B: HOLOGRAPHIC FORM FACTORS

This Appendix includes a summary of how 5D gauge fields and fermions are treated holographically, and how their form factors are derived. The 5D scenarios that we present will be simplified versions of the full setup considered in the main text, however the results arrived at will be used to build the form factors for the $SO(6)/SO(5)$ model. For a full review of holographic techniques on gauge fields and fermion fields we refer the reader to [38,47], respectively.

### 1. Gauge fields

To quadratic order, the action for a non-Abelian 5D gauge field can be written as

$$S = \frac{1}{2g_5^2} \int d^5x \sqrt{|g|} \left[ -\frac{1}{2} F^{\mu\nu,A} F_{\mu\nu}^A - F^{\mu5,A} F_{\mu\nu}^A \right] \quad (B1)$$

with $g$ being the determinant of the metric, and $A$ labeling the generators of the bulk gauge field. The IR boundary conditions for the unbroken ($A = a$) and broken ($A = \hat{a}$) generators are Neumann and Dirichlet, respectively, i.e.,

$$\partial_z A_\mu^a(p, z)|_{z=R'} = 0, \qquad A_5^a(p, R') = 0$$
$$\partial_z A_5^{\hat{a}}(p, z)|_{z=R'} = 0, \qquad A_\mu^{\hat{a}}(p, R') = 0. \quad (B2)$$

Throughout the paper we refer to Neumann and Dirichlet boundary conditions with a "+" and a "−," respectively. The UV boundary conditions are used to define a source field, i.e., the 4D degree of freedom with which we will define the effective theory. For the unbroken and broken generators these boundary conditions are

$$A_\mu^a(p, R) \equiv A_\mu(p), \qquad A_\mu^{\hat{a}}(p, R) \equiv 0. \quad (B3)$$

In the Kaluza-Klein method of treating 5D gauge fields this is equivalent to having Neumann and Dirichlet boundary conditions on the UV brane for the unbroken and broken generators, respectively. This in turn implies Dirichlet and Neumann UV boundary conditions for the $A_5^a$ and $A_5^{\hat{a}}$ components, respectively, and thus massless modes in the spectrum for the $A_\mu^a$ and $A_5^{\hat{a}}$ fields. If we want to impose Dirichlet UV boundary conditions for any of the fields for which we do define a source field we can simply introduce a Higgs mechanism resulting in a large mass term for that field to the UV brane.

Solving the bulk 5D equations of motion for the $A_{\mu,5}^A$ fields, and inserting these back into the action, allows one to obtain the holographic action. For the gauge fields the effective action is found to be

$$S_{\text{hol}_g} = -\frac{1}{2g_5^2} P_T^{\mu\nu} (A_\mu^a \Pi_+(p^2) A_\nu^a + A_\mu^{\hat{a}} \Pi_-(p^2) A_\nu^{\hat{a}}) \quad (B4)$$

with $P_T^{\mu\nu} = \eta^{\mu\nu} - \frac{p^\mu p^\nu}{p^2}$. The form factors are calculated to be

$$\Pi_-(p_E^2) = p_E \frac{K_1(p_E R')I_0(p_E R) + I_1(p_E R')K_0(p_E R)}{K_1(p_E R')I_1(p_E R) - I_1(p_E R')K_1(p_E R)}$$

$$\Pi_+(p_E^2) = p_E \frac{K_1(p_E R')I_0(p_E R) - I_1(p_E R')K_0(p_E R)}{K_0(p_E R')I_1(p_E R) + I_0(p_E R')K_1(p_E R)} \quad (B5)$$

where $p_E$ is the Wick rotated momentum. In their original Minkowski space form these form factors have zeros and poles which will be used to determine the mass spectra of the 4D eigenstates in the theory. We call these mass eigenstates Kaluza-Klein modes. With these form factors we can write down all the form factors required in the gauge sector of the $SO(6)/SO(5)$ model.

### 2. Fermion fields

Take two 5D Dirac fermions living in the bulk of the RS model, $\Psi_1$ and $\Psi_2$. The UV boundary conditions for these fields choose the $\Psi_{1L}$ and $\Psi_{2R}$ Weyl components as the dynamical source fields, i.e., $\Psi_1(p, R) \equiv \Psi_{1L}(p)$ and $\Psi_2(p, R) \equiv \Psi_{2R}(p)$. On the IR brane the boundary conditions are determined by dimensionless IR mass mixings ($\tilde{m}$) between the two 5D fermions. The action for such a scenario can be written as

$$\mathcal{L} = \int dx^5 \sqrt{|g|} \left\{ \frac{i}{2} \bar{\Psi}_1 \gamma^M \partial_M \Psi_1 - \frac{i}{2} (\partial_M \bar{\Psi}_1) \gamma^M \Psi_1 \right.$$
$$- m_1 \bar{\Psi}_1 \Psi_1 \frac{i}{2} \bar{\Psi}_2 \gamma^M \partial_M \Psi_2 - \frac{i}{2} (\partial_M \bar{\Psi}_2) \gamma^M \Psi_2 - m_2 \bar{\Psi}_2 \Psi_2$$
$$- \delta(z - R') \tilde{m}(\bar{\Psi}_{1L} \Psi_{2R} + \bar{\Psi}_{2R} \Psi_{1L})$$
$$\times \delta(z - R) \frac{1}{2} (\bar{\Psi}_{1L} \Psi_{1R} - \bar{\Psi}_{2L} \Psi_{2R})$$
$$\left. - \delta(z - R') \frac{1}{2} (\bar{\Psi}_{1L} \Psi_{1R} - \bar{\Psi}_{2L} \Psi_{2R}) \right\}. \quad (B6)$$

The terms on the last line are necessary additions in order to satisfy the boundary conditions. The IR boundary conditions following from this are

$$\Psi_{1R}(R') = -\tilde{m}\Psi_{2R}(R'), \qquad \Psi_{2L}(R') = \tilde{m}\Psi_{1L}(R'). \quad (B7)$$

In the Kaluza-Klein picture, choosing a LH (RH) source field on the UV brane corresponds to a Neumann boundary condition for the LH (RH) component, with a Dirichlet boundary condition for the other chirality. Taking the $\tilde{m} \to 0$ limit on the IR brane we obtain Dirichlet boundary conditions for the $\Psi_{1R}$ and $\Psi_{2L}$ components, which implies Neumann boundary conditions for the $\Psi_{1L}$ and $\Psi_{2R}$ components. Therefore there exists massless zero modes for $\Psi_{1L}$ and $\Psi_{2R}$ in the spectrum, with their localization in the extra dimension determined by the 5D mass parameter $m_i$. When $\tilde{m} \neq 0$ we still have two massless modes in the model, except now these modes are an admixture of $\Psi_1$ and $\Psi_2$.

We solve the equations of motion for the 5D fermions such that $\Psi(p,z) \sim G(p,z)\Psi(p)$, where $G(p,z)$ is some holographic profile and $\Psi(p)$ is the holographic source field defined by the UV boundary condition. The UV and IR boundary conditions are satisfied by fixing integration constants in $G(p,z)$. With these holographic profiles the bulk dynamics can be integrated out and we obtain the following effective action for the source fields,

$$\mathcal{L} = \bar{\Psi}_{1L}\,\slashed{p}\,\Pi_1^f(p)\Psi_{1L} + \bar{\Psi}_{2R}\,\slashed{p}\,\Pi_2^f(p)\Psi_{2R} - \bar{\Psi}_{1L}M^f(p)\Psi_{2R} \tag{B8}$$

where the form factors encode the mass spectra and mass mixings of the fields, analogously to those calculated in the case of gauge fields. In terms of the 5D parameters these can be expressed as

$$\Pi_1^f(p, c_1, c_2, \tilde{m}) = \frac{1}{p} \frac{G_p^+(-c_2)G_p^-(c_1) + \tilde{m}^2 G_p^-(c_2)G_p^+(-c_1)}{G_p^+(c_1)G_p^+(-c_2) - \tilde{m}^2 G_p^-(-c_1)G_p^-(c_2)}$$

$$\Pi_2^f(p, c_1, c_2, \tilde{m}) = \frac{1}{p} \frac{G_p^-(-c_2)G_p^+(c_1) + \tilde{m}^2 G_p^+(c_2)G_p^-(-c_1)}{G_p^+(c_1)G_p^+(-c_2) - \tilde{m}^2 G_p^-(-c_1)G_p^-(c_2)}$$

$$M^f(p, c_1, c_2, \tilde{m}) = \frac{\tilde{m}}{2} \cdot \frac{G_p^-(-c_2)G_p^+(c_2) + G_p^-(-c_2)G_p^-(c_2) + G_p^+(-c_1)G_p^+(c_1) + G_p^-(-c_1)G_p^-(c_1)}{G_p^+(c_1)G_p^+(-c_2) - \tilde{m}^2 G_p^-(-c_1)G_p^-(c_2)} \tag{B9}$$

where the $G_p^\pm$ are the 5D holographic functions derived from the equations of motion in the bulk, and the parameters $c_{1,2}$ are the dimensionless mass parameters defined by $c_i = m_i R$. After a Wick rotation these holographic functions can be written as

$$G_p^+(p_E, c, z) = -\frac{2i}{\pi}\sqrt{r}\left(K_{c-\frac{1}{2}}(p_E R')I_{c+\frac{1}{2}}(p_E z) + I_{c-\frac{1}{2}}(p_E R')K_{c+\frac{1}{2}}(p_E z)\right)$$

$$G_p^-(p_E, c, z) = -\frac{2}{\pi}\sqrt{r}\left(K_{c-\frac{1}{2}}(p_E R')I_{c-\frac{1}{2}}(p_E z) - I_{c-\frac{1}{2}}(p_E R')K_{c-\frac{1}{2}}(p_E z)\right). \tag{B10}$$

We present the Wick rotated result because this is what we will use in calculating the Higgs potential, however it is trivial to obtain the original result with $p_E \to -ip$.